\documentclass[prd,aps,twocolumn,nofootinbib,preprintnumbers]{revtex4}
\usepackage{graphicx,amsmath}
\usepackage{comment}
\usepackage{color}
\begin{document}
%%%%%%%%%%%%%%%%%%%%%%%%%%%%%%%%%%%%%%%%%%%%%%%%%%

\newcommand{\mtrem}[1]{{\color{red} \bf $[[$ MT: #1 $]]$}}
\newcommand{\Jrem}[1]{{\color{blue} \bf $[[$ RJ: #1 $]]$}}

\def\beq{\begin{eqnarray}}
\def\eeq{\end{eqnarray}}
\newcommand{\gsim}{ \mathop{}_{\textstyle \sim}^{\textstyle >} }
\newcommand{\lsim}{ \mathop{}_{\textstyle \sim}^{\textstyle <} }
\newcommand{\vev}[1]{ \left\langle {#1} \right\rangle }
\newcommand{\bra}[1]{ \langle {#1} | }
\newcommand{\ket}[1]{ | {#1} \rangle }
\newcommand{\EV}{ {\rm eV} }
\newcommand{\KEV}{ {\rm keV} }
\newcommand{\MEV}{ {\rm MeV} }
\newcommand{\GEV}{ {\rm GeV} }
\newcommand{\TEV}{ {\rm TeV} }
\newcommand{\bea}{\begin{eqnarray}}   
\newcommand{\eea}{\end{eqnarray}}
\newcommand{\bear}{\begin{array}}  
\newcommand {\eear}{\end{array}}
\newcommand{\bef}{\begin{figure}}  
\newcommand {\eef}{\end{figure}}
\newcommand{\bec}{\begin{center}}  
\newcommand {\eec}{\end{center}}
\newcommand{\non}{\nonumber}  
\newcommand {\eqn}[1]{\beq {#1}\eeq}
\newcommand{\la}{\left\langle}  
\newcommand{\ra}{\right\rangle}
\newcommand{\ds}{\displaystyle}
\def\SEC#1{Sec.~\ref{#1}}
\def\FIG#1{Fig.~\ref{#1}}
\def\EQ#1{Eq.~(\ref{#1})}
\def\EQS#1{Eqs.~(\ref{#1})}
\def\GEV#1{10^{#1}{\rm\,GeV}}
\def\MEV#1{10^{#1}{\rm\,MeV}}
\def\KEV#1{10^{#1}{\rm\,keV}}
\def\lrf#1#2{ \left(\frac{#1}{#2}\right)}
\def\lrfp#1#2#3{ \left(\frac{#1}{#2} \right)^{#3}}

%%%%%%%%%%%%%%%%%%%%%%%%%%%%%%%%%%%%%%%%%%%%%%%%%%

\preprint{KEK-TH-1896}
\title{Probing classically conformal $B-L$ model by gravitational waves}
\renewcommand{\thefootnote}{\alph{footnote}}

\author{
Ryusuke Jinno,
and
Masahiro Takimoto}

\affiliation{
Theory Center, High Energy Accelerator Research Organization (KEK), 1-1 Oho, Tsukuba, Ibaraki 305-0801, Japan\\
}

\begin{abstract}

We study the cosmological history of the classical conformal $B-L$ gauge extension of
the standard model,
in which the physical scales are generated via the Coleman-Weinberg-type symmetry breaking. 
Especially, we consider the thermal phase transition of the U$(1)_{B-L}$ symmetry in the early universe
and resulting gravitational-wave production.
Due to the classical conformal invariance, 
the phase transition tends to be a first-order one with ultra-supercooling,
which enhances the strength of the produced gravitational waves. 
We show that, 
requiring 
(1) U$(1)_{B-L}$ is broken after the reheating, 
(2) the $B-L$ gauge coupling does not blow up below the Planck scale, 
(3) the thermal phase transition completes in almost all the patches in the universe, 
the gravitational wave spectrum can be as large as $\Omega_{\rm GW} \sim 10^{-8}$ 
at the frequency $f \sim 0.01$--$1$Hz for some model parameters,
and a vast parameter region can be tested by future interferometer experiments
such as eLISA, LISA, BBO and DECIGO.

\end{abstract}

\maketitle

%%%%%%%%%%%%%%%%%%%%%%%%%%%%%%%%%%%%%%%%%%%%%%%%%%
\section{Introduction}
%%%%%%%%%%%%%%%%%%%%%%%%%%%%%%%%%%%%%%%%%%%%%%%%%%

Detection of the gravitational waves (GWs) is one of the most promising tools to probe 
the early Universe as well as astrophysical dynamics. 
Possible GW sources in the early Universe include for example
inflationary quantum fluctuations~\cite{Starobinsky:1979ty}, 
preheating~\cite{Khlebnikov:1997di},
cosmic strings~\cite{Vilenkin:2000jqa}
and
cosmic phase transitions~\cite{Witten:1984rs,Hogan:1986qda}.
Especially, the last possibility has been extensively studied 
in the context of the electroweak transition triggered by the standard model (SM)
Higgs field.
Though the electroweak transition within the SM
was found to be too weak to produce an observable amplitude of GWs~\cite{Rummukainen:1998as},
various extensions of the SM predict first order phase transitions 
with a large amplitude of GWs~\cite{Espinosa:2008kw,Ashoorioon:2009nf,Das:2009ue,Sagunski:2012pzo,Kakizaki:2015wua,Jinno:2015doa,Apreda:2001tj,Apreda:2001us,Jaeckel:2016jlh,Huber:2015znp,Leitao:2015fmj,Huang:2016odd,Dev:2016feu}.

In considering physics beyond the SM, 
the Higgs sector may provide us some important clue.
Especially, the smallness of the electroweak scale, 
compared to the 
Planck scale or grand unification scale (if it is realized in nature), 
has long been considered as unnatural,
and puzzled people as ``naturalness problem" or
more specifically ``gauge hierarchy problem".
One of the most popular ways to solve this problem is to impose supersymmetry (SUSY) on the theory.
This additional symmetry introduces so-called superpartners with opposite statistics 
to each particle content in the SM, 
and the Higgs mass is protected from the quadratic divergence
because of the cancellation with bosonic and fermionic loops.
As long as the soft-breaking scale of SUSY is close to the electroweak scale,
this provides a convincing solution to the naturalness problem.
However, despite all the efforts made by the high-energy community,
the current data at Large Hadron Collider (LHC) suggest no traces of superpartners.
Therefore, while SUSY remains to be an attractive solution to the naturalness problem,
this situation leads us to look for other possibilities.

Here let us recall the argument by Bardeen~\cite{Bardeen:1995kv} 
that the SM itself has no gauge hierarchy problem as long as we consider perturbative region.
This is because the SM has an approximate scale symmetry which is broken only logarithmically, 
and therefore the seemingly quadratic dependence on the cutoff regulator $\Lambda$
must be canceled out in the Higgs mass.
No fine-tuning is required in this cancellation, 
because it is protected from this quadratic divergence by the approximate scale symmetry.
The gauge hierarchy problem appears only when 
nonperturbative effects such as Landau pole appear, 
or when we consider some UV completion of the theory.
Whether the Higgs mass is protected by the approximate scale symmetry 
when we take these into account
depends on the formulation of the theory in the UV.

Given this, it might be interesting to consider the possibility 
that the UV completion is such that it leaves no tree masses in its low energy effective action, 
rather than leaving a small portion as the EW scale.
In such cases, all the mass scales must be generated by dimensional transmutation
such as Coleman-Weinberg mechanism~\cite{Coleman:1973jx}.
Though it was found that this mechanism cannot be applied directly to the SM Higgs sector
since the predicted Higgs mass $\sim 10$ GeV is experimentally excluded,
phenomenologically viable models still exist 
if one includes an additional scalar to the model 
so that the mass scale comes from the breaking of the conformal invariance of that field~\cite{Hempfling:1996ht,Chang:2007ki,Foot:2007as,Meissner:2006zh,Iso:2009ss,Iso:2009nw}
(see also \cite{Helmboldt:2016mpi} for \cite{Meissner:2006zh}).
In Ref.~\cite{Meissner:2006zh}, 
Meissner and Nicoli called 
the property free from tree-level masses 
(though somewhat misleading) ``classically conformal".  
  
In this paper, we study the cosmological history 
of the classically conformal $B-L$ model proposed in Refs.~\cite{Iso:2009ss,Iso:2009nw}.
Especially we are interested in GW production during the phase transition of 
the $B-L$ breaking field.
As we see later, the ``classical conformal" property is the key to the large energy density
released during the phase transition and the resulting large amplitude of GWs,
and these features are considered to be shared in many classical conformal models.
Since now ground-based detectors such as 
KAGRA~\cite{Somiya:2011np},
VIRGO~\cite{TheVirgo:2014hva} and Advanced LIGO~\cite{Harry:2010zz}
are in operation,
and space interferometer experiments 
eLISA~\cite{Seoane:2013qna}, Big-Bang Observer (BBO)~\cite{Harry:2006fi} 
and DECi-hertz Interferometer Observatory (DECIGO)~\cite{Seto:2001qf}
have been proposed,
it would be interesting to study the GW production in these classically conformal models.

The organization of the paper is as follows.
In Sec.~\ref{sec_setup} we first write down the model and see its finite-temperature behavior.
Then in Sec.~\ref{sec_history} we study the cosmic history it follows.
Since GW production occurs at the time of phase transition,
we elaborate the criteria for the transition in this section.
In Sec.~\ref{sec_GW} we discuss GW production in this model
and study the detectability of these GWs by future interferometer experiments.
We summarize in Sec.~\ref{sec_conclusions}.

%%%%%%%%%%%%%%%%%%%%%%%%%%%%%%%%%%%%%%%%%%%%%%%%%%
\section{Setup}
\label{sec_setup}
%%%%%%%%%%%%%%%%%%%%%%%%%%%%%%%%%%%%%%%%%%%%%%%%%%

In this section we first write down the setup of the model, 
and then see the finite-temperature behavior of the effective potential.

%%%%%%%%%%%%%%%%%%%%%%%%%%%%%%%%%%%%%%%%%%%%%%%%%%
\subsection{Model}
%%%%%%%%%%%%%%%%%%%%%%%%%%%%%%%%%%%%%%%%%%%%%%%%%%

The minimal $B-L$ extension of the Standard Model with classical conformal symmetry,
as discussed in Refs.~\cite{Iso:2009ss,Iso:2009nw}, 
is based on the argument that 
once the classical conformal invariance and its violation by quantum anomalies are imposed on SM,
the model becomes free from the gauge hierarchy problem~\cite{Bardeen:1995kv}.
This model has the gauge symmetry 
SU$(3)_c\times$SU$(2)_L\times$U$(1)_Y\times$U$(1)_{B-L}$,
and introduces three generations of right handed neutrinos $\nu_R^i$ ($i = 1,2,3$).
In addition, it contains a complex scalar field $\Phi$, 
in order to break the U$(1)_{B-L}$ gauge symmetry by the vacuum expectation value (VEV) and 
to induce the masses of right handed neutrinos.
The matter contents of the model are listed in Table~\ref{table_matcon}.

%%%%%%%%%%
\begin{table}[t]
\begin{tabular}{|c||c|c|c|c|}
\hline
&SU$(3)_c$&SU$(2)_L$&U$(1)_Y$&U$(1)_{B-L}$ \\ \hline\hline
$q_L^i$& {\bf{3}}&{\bf 2}&+1/6&+1/3\\
$u_R^i$& {\bf{3}}&{\bf 1}&+2/3&+1/3\\
$d_R^i$& {\bf{3}}&{\bf 1}&-1/3&+1/3\\ \hline
$l_L^i$& {\bf{1}}&{\bf 2}&+1/6&-1\\
$e_R^i$& {\bf{1}}&{\bf 1}&-1&-1\\
$\nu_R^i$& {\bf{1}}&{\bf 1}&0&-1\\ \hline
$H$& {\bf{1}}&{\bf 2}&-1/2&0\\
$\Phi$& {\bf{1}}&{\bf 1}&0&+2\\ \hline
\end{tabular}
\caption{Matter contents of the classically conformal $B-L$ model. 
In addition to the standard model matters, three generations of right-handed neutrinos $\nu_R^i$ and
a $B-L$ charged complex scalar field $\Phi$ are introduced.}
\label{table_matcon}
\end{table}
%%%%%%%%%%

This model has the following Yukawa interactions in addition to the SM
\begin{align}
\mathcal{L}_Y
&\supset -Y_D^{ij}\bar{\nu}_R^i H^\dagger l_L^j-\frac{1}{2}Y_M^{i}\Phi\bar{\nu}_R^{ic}\nu_R^i+\text{h.c.},
\end{align}
where we can assume the Yukawa coupling $Y_M^i$ to have a diagonal form without loss of generality.
Neutrino masses are generated by the seesaw mechanism~\cite{seesaw}
through the VEVs of the standard model Higgs boson $H$ and $\Phi$.
On the other hand, 
the scalar potential of this model consists of the following terms
\begin{align}
V
&= \lambda_H|H|^4+\lambda|\Phi|^4-\lambda'|\Phi|^2|H|^2,
\end{align}
where only four-point couplings appear due to the assumption of the classical conformal invariance.
The additional scalar field $\Phi$ can obtain a VEV $M\equiv \sqrt{2}\langle \Phi\rangle$
through the running of the coupling $\lambda$.  

Here we mention the viable parameter range.
The VEV $M$ is bounded from below by the constraint on the mass of $B-L$ gauge boson $Z'$
through the relation $m_{Z'} = 2g_{B-L}M$, with $g_{B-L}$ being the $B-L$ gauge coupling constant.
The current constraint reads $M\gtrsim 10$ TeV~\cite{Okada:2016gsh}.
On the other hand, if $M$ is much larger than the electroweak scale, 
the Higgs mass term obtains sizable corrections proportional to $M^2$
through loop diagrams~\cite{Iso:2009nw}.
Therefore we expect that the value of $M$ is not far from the electroweak scale, 
since otherwise the tuning becomes more and more severe.
In this paper we focus on $10^3$ GeV $\lesssim M \lesssim$ $10^9$ GeV.
In order to realize the electroweak vacuum, the coupling $\lambda'$ must be somewhat suppressed and
we neglect the coupling $\lambda'$ below.
We also neglect the Yukawa couplings $Y_{D/M}$ by assuming $g_{B-L}\gtrsim Y_{D/M}$
for simplicity.

The zero-temperature effective potential for $\phi \equiv \sqrt{2}\mathcal{R}[\Phi]$
at one-loop level is written as~\cite{Meissner:2008uw,Iso:2009nw} 
\begin{align}
V_0(\phi,t)
&= \frac{1}{4}\lambda(t)G^4(t)\phi^4,
\end{align}
where $t = \log(\phi/\mu)$ with $\mu$ being the renormalization scale and
\begin{align}
G(t)
&= \exp\left[-\int_0^tdt'\gamma(t')\right],
\;\;\;
\gamma(t)
= -\frac{a_2}{32\pi^2}g_{B-L}(t)^2,
\end{align}
with $a_2=24$.
The gauge and self coupling strength $\alpha_{B-L}\equiv g_{B-L}^2/4\pi$ and 
$\alpha_\lambda \equiv \lambda/4\pi$
obey the following renormalization group equations
\begin{align}
&2\pi \frac{d\alpha_{B-L}(t)}{dt}
= b\alpha_{B-L}(t)^2, \\
&2\pi \frac{d\alpha_\lambda(t)}{dt}
= a_1\alpha_\lambda(t)^2+8\pi\alpha_\lambda(t) \gamma(t)+a_3\alpha_{B-L}(t)^2,
\end{align}
with $b=12$, $a_1=10$ and $a_3=48$.
In the following we take the renormalization scale $\mu$ to be $M$.
This allows one to rewrite the condition 
$\frac{dV}{d\phi}\big|_{\phi=M} = 0$ as
\begin{align}
a_1\alpha_\lambda(0)^2 + a_3\alpha_{B-L}(0)^2 + 8\pi\alpha_\lambda(0)
&= 0,
\label{eq_cond_vev}
\end{align}
which means that $\alpha_{\lambda}(0)$ is determined by $\alpha_{B-L}(0)$.
Thus, the scalar sector in our setup has only two parameters, $M$ and $\alpha_{B-L}(0)$.
In the parameter region we are interested in, 
the first term is neglected in Eq.~(\ref{eq_cond_vev})
and therefore $-\alpha_\lambda(0) \sim \alpha_{B-L}(0)^2 \ll 1$ holds.
With the help of Eq.~(\ref{eq_cond_vev}), we obtain the mass relation 
between the masses of $\phi$ and $Z'$
\begin{align}
\left( \frac{m_\phi}{m_{Z'}} \right)^2
&\simeq  \frac{6}{\pi} \alpha_{B-L}(0).
\end{align}

The running of the couplings can be solved analytically, 
and the scalar potential is given by~\cite{Iso:2009ss}
\begin{align}
V_0(\phi,t)
&=\frac{\pi\alpha_\lambda(t)}{\left(1-\frac{b}{2\pi}\alpha_{B-L}(0)t\right)^{a_2/b}} \phi^4,\
\label{eq_V0}
\end{align}
where
\begin{align}
&\alpha_{B-L}(t)
= \frac{\alpha_{B-L}(0)}{1-\frac{b}{2\pi}\alpha_{B-L}(0)t},
\label{eq_aBLSolution}
\\
&\alpha_\lambda(t)
= \frac{a_2+b}{2a_1}\alpha_{B-L}(t) \nonumber 
\\
&\;\;\;\;\;\;\;\;\;\;\;\;\;
+\frac{A}{a_1}\alpha_{B-L}(t)\tan\left[
\frac{A}{b}\ln\left[\alpha_{B-L}(t)/\pi \right]+C
\right].
\end{align}
Here $A$ is defined as $A \equiv \sqrt{a_1a_3-(a_2+b)^2/4}$,
and the coefficient $C$ is determined so that Eq.~(\ref{eq_cond_vev}) holds.

%%%%%%%%%%%%%%%%%%%%%%%%%%%%%%%%%%%%%%%%%%%%%%%%%%
\subsection{Finite temperature effective potential}
%%%%%%%%%%%%%%%%%%%%%%%%%%%%%%%%%%%%%%%%%%%%%%%%%%

In order to follow the dynamics of the scalar field $\Phi$ in the early universe, 
we must take into account the finite-temperature effect on the effective potential.
Since the system we consider has two typical scales $\phi$ and $T$ with $T$ being the temperature of the universe,
we define the renormalization scale parameter $u$ instead of $t$ as
\begin{align}
u
&\equiv \log(\Lambda/M),
\label{eq_uDef}
\end{align}
where
\begin{align}	
\Lambda
&\equiv
\max (\phi,T).
\end{align}
Note that $\Lambda$ represents the typical scale of the system.
Then, the one-loop level effective potential can be written as
\begin{align}
V_{\rm eff}(\phi,T)
&= V_0(\phi,u)+V_T(\phi,T).
\label{eq_effpot}
\end{align}
Here $V_0$ indicates the zero-temperature potential (\ref{eq_V0}),
while $V_T$ denotes the thermal potential
\begin{align}
V_T(\phi,T)
&=\frac{3}{2}V_T^B(m_V(\phi)/T,T)+V_{\rm daisy}(\phi,T),
\end{align}
where
\begin{align}
V_T^B(x,T)
&\equiv \frac{T^4}{\pi^2}\int_0^\infty dz~z^2\ln \left[ 1-e^{-\sqrt{z^2+x^2}} \right],
\end{align}
is the bosonic one-loop contribution, and
\begin{align}
V_{\rm daisy}(\phi,T)
&= -\frac{T}{12\pi}\left[m_V^3(\phi,T) - m_V^3(\phi)\right],
\end{align}
is so-called daisy subtraction~\cite{Arnold:1992rz}.
In the above expressions, the masses of the gauge boson are given by
\begin{align}
	m_V(\phi)&=2g_{B-L}(t)\phi,\\
	m_V^2(\phi,T)&=m_V^2(\phi)+c_t g_{B-L}^2(t)T^2,~~c_t=4.
\end{align}
Note that we have neglected the contribution from $\phi$'s self-interaction
to the thermal potential, since it is much smaller than the one from the gauge interaction.

%%%%%%%%%%%%%%%%%%%%%%%%%%%%%%%%%%%%%%%%%%%%%%%%%%
\section{History of the universe}
\label{sec_history}
%%%%%%%%%%%%%%%%%%%%%%%%%%%%%%%%%%%%%%%%%%%%%%%%%%

%%%%%%%%%%%%%%%%%%%%%%%%%%%%%%%%%%%%%%%%%%%%%%%%%%
\subsection{Overview of our scenario}
%%%%%%%%%%%%%%%%%%%%%%%%%%%%%%%%%%%%%%%%%%%%%%%%%%

\begin{figure}
\begin{center}
\includegraphics[scale=0.85]{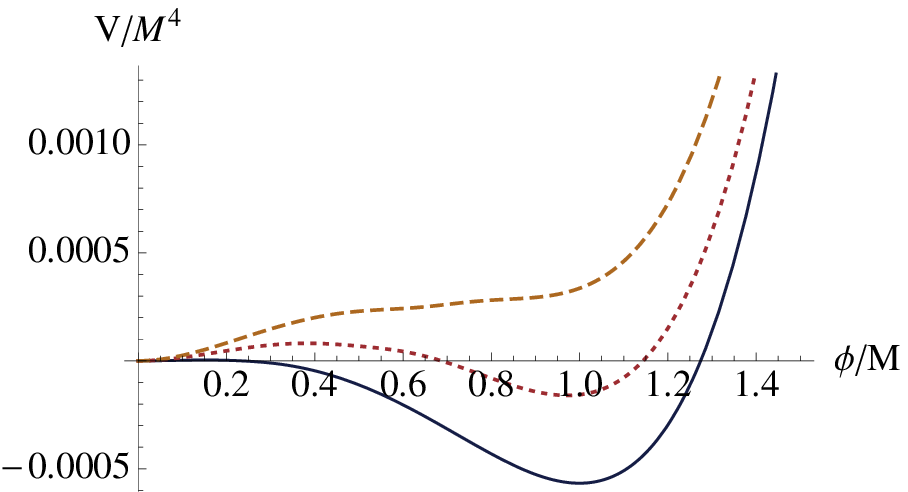}
\caption {\small
Plot of the finite-temperature effective potential $V_{\rm eff}$.
Parameters are taken to be 
$\alpha_{B-L}(0) = 0.01$ and $T/M = 0.1$ (blue), $0.2$ (red) and $0.25$ (yellow).
The origin is the true vacuum for high enough temperature, while it is a false one 
after the temperature drops.
}
\label{fig_pot}
\end{center}
\begin{center}
\includegraphics[scale=0.85]{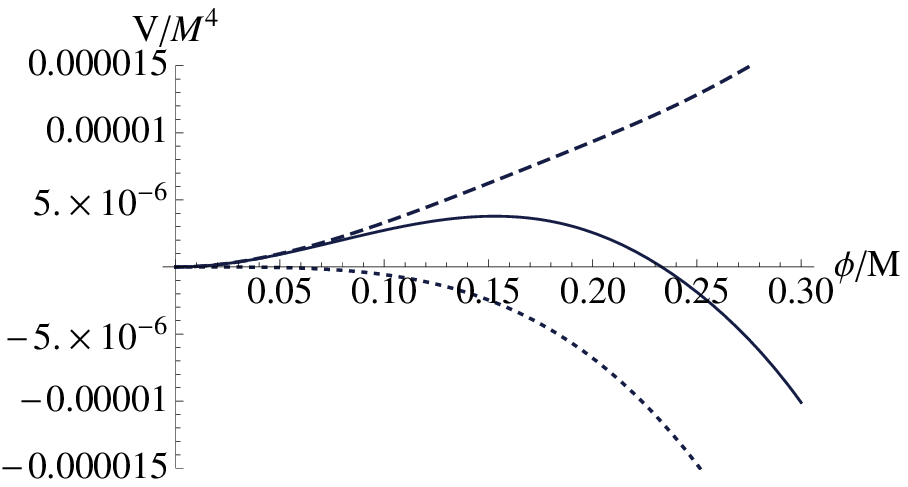}
\caption {\small
Blow-up of Fig.~\ref{fig_pot}.
Solid line is the same as in the blue line of Fig.~\ref{fig_pot},
while dashed and dotted lines show the contributions from the thermal potential
$V_T$ and the zero-temperature potential $V_0$, respectively.
Because of the classical conformal requirement,
the thermal contribution dominates the zero-temperature one near the origin
(see Eq.~(\ref{eq_Veff_approx})).
}
\label{fig_pot_blowup}
\end{center}
\end{figure}

Here we briefly give the overview of the cosmological scenario realized in the present model.
For that purpose it would be helpful to approximate the effective potential as
\begin{align}
V_{\rm eff}
&\sim \frac{g_{B-L}^2(u)}{2}T^2\phi^2 + \frac{\lambda_{\rm eff}(u)}{4}\phi^4,
\label{eq_Veff_approx}
\end{align}
with
$\lambda_{\rm eff}(u) \equiv 4\pi\alpha_\lambda(u) / \left( 1-\frac{b}{2\pi}\alpha_{B-L}(0)u \right)^{a_2/b}$
(see Eq.~(\ref{eq_V0})).
If the temperature is high enough $T \gg M$, 
the effective potential has the unique minimum at $\phi = 0$ 
since the self-coupling satisfies $\lambda_{\rm eff} > 0$.
On the other hand, for $T \ll M$, 
the self coupling around $\phi \lesssim T$ becomes negative
and the origin $\phi=0$ is made to be a false vacuum.
Fig.~\ref{fig_pot} shows the temperature dependence of $V_{\rm eff}$.
One sees that $\phi = 0$ is the true vacuum at high temperature, while it becomes a false one as the temperature decreases.
Note that $\phi = 0$ continues to be a local minimum, as is understood from Fig.~\ref{fig_pot_blowup},
since the contribution from the first term in the RHS of Eq.~(\ref{eq_Veff_approx})
always dominates the one from the second term for $\phi$ near the origin.

With this behavior of the effective potential,
the evolution of the universe is summarized as follows.
Assuming that the reheating temperature is so high that
U$(1)_{B-L}$ is restored at the time of the reheating, 
$\phi$ is first trapped at the origin of the effective potential.
After the temperature drops to $T = T_c \sim M$,
the origin becomes a false vacuum and 
for some cases
the universe experiences a first-order phase transition 
associated with the tunneling of the $\phi$ field.
This transition triggers bubble production and subsequent GW production, 
as we see in Sec.~\ref{sec_GW}.
This makes the most interesting part of our scenario.

However, note that 
the universe does not necessarily experience a first-order phase transition
even if the origin of $\phi$ becomes a false vacuum.
This is because the transition rate must exceed the expansion rate of the universe in order to complete the transition.
Therefore we have to provide some criteria for the transition, 
which we explain in Sec.~\ref{subsec_criteria}.
If this criteria is satisfied, 
a given spacial point in the false vacuum well before the typical transition time 
finds itself in the true vacuum with a probability close to unity in the infinite future.
In addition, it finds itself surrounded by a continuum of true vacuum which well exceeds 
the horizon size covered by CMB observations.

%%%%%%%%%%%%%%%%
%%%%%%%%%%%%%%%%

%%%%%%%%%%%%%%%%%%%%%%%%%%%%%%%%%%%%%%%%%%%%%%%%%%
\subsection{Criteria for the phase transition}
\label{subsec_criteria}
%%%%%%%%%%%%%%%%%%%%%%%%%%%%%%%%%%%%%%%%%%%%%%%%%%

%%%%%%%%%%%%%%%%%%%%%%%%%%%%%%%%%%%%%%%%%%%%%%%%%%
\subsubsection{Nucleation rate}
%%%%%%%%%%%%%%%%%%%%%%%%%%%%%%%%%%%%%%%%%%%%%%%%%%

In finite temperature field theory, the nucleation rate per unit volume $\Gamma$ is 
given by~\cite{Linde:1977mm,Linde:1981zj}
\begin{align}
&\Gamma (T)
= A(T) e^{-S_3(T)/T},
\label{eq_Gamma1}
\end{align}
with
\begin{align}
&S_3(T)
= \int d^3x \left[ \frac{1}{2} (\nabla \phi)^2 + \left( V_{\rm eff}(\phi,T) - V_{\rm eff}(0,T) \right) \right],
\label{eq_S3}
\end{align}
where $A$ denotes a prefactor which is typically of $\mathcal{O}(T^4)$\footnote{
We consider $g_{B-L}\lesssim 0.3$ region in the following. 
In such a case, the effects from the prefactor are negligible 
(see Eqs. (8) and (9) in~\cite{Strumia:1999fv}).
We have checked that the nucleation rate of the $O$(4) symmetric vacuum bubble is negligible.
}.
The configuration of $\phi$ in $S_3$ is 
estimated from saddle point approximation of the path integral 
with $O(3)$ symmetry assumption, and is obtained from the following equation
\begin{align}
\frac{d^2 \phi}{dr^2} + \frac{2}{r} \frac{d\phi}{dr} - \frac{\partial V_{\rm eff}}{\partial \phi}
&= 0.
\end{align}
Here the boundary conditions are taken to be
\begin{align}
\phi(r = \infty)
&= 
\phi_{\rm false}, 
\;\;\;\;\;\;
\frac{d\phi}{dr} (r=0)
= 0.
\end{align}
Here we rewrite the transition rate (\ref{eq_Gamma1}) as
\begin{align}
\Gamma(T)
&=
Be^{-S(T)}, 
\label{eq_Gamma2} 
\end{align}
with
\begin{align}
B
&\equiv M^4 \frac{A(T)}{T^4}, \\
S(T)
&\equiv \frac{S_3(T)}{T} - 4\log(T/M).
\label{eq_S}
\end{align}
Since $A$ is of ${\mathcal O}(T^4)$ and the transition rate is 
dominantly determined by $S$, we simply set $B$ to be $M^4$ below.
The purpose of the redefinition (\ref{eq_Gamma2}) is to 
make the $T$-dependence of the nucleation rate $\Gamma$ 
to appear in the exponent $S(T)$\footnote{ 
In the literature, only the $T$-dependence of $S_3/T$ is often discussed
and that of $A(T)$ is neglected,
because the former affects $\Gamma$ much more strongly.
However,  
the $T$-dependence of $S_3/T$ is quite small in our model as we see later,
and therefore this procedure is necessary to estimate the behavior of the 
transition rate more correctly.
}.

%%%%%%%%%%%%%%%%%%%%%%%%%%%%%%%%%%%%%%%%%%%%%%%%%%
\subsubsection{False vacuum probability}
%%%%%%%%%%%%%%%%%%%%%%%%%%%%%%%%%%%%%%%%%%%%%%%%%%

We introduce the false vacuum probability $p(t)$, which is defined as the probability for a given spacial point
in the false vacuum well before the typical transition time to remain in the false vacuum at time $t$~\cite{Turner:1992tz}.
Assuming the de Sitter spacetime background\footnote{
In our setup, 
the transition mainly occurs after the scalar potential dominates the universe.
In this case the assumption of de Sitter background gives a good approximation.
Though the transition occurs before the scalar domination for some parameters,
Eq.~(\ref{eq_p}) still gives a good approximation in this case
since the transition finishes instantaneously compared to the Hubble time.
}, 
it is written as
\begin{align}
p(t)
&=e^{-I(t)},
\label{eq_p}
\end{align}
with
\begin{align}
I(t)
&=\frac{4\pi}{3}\int_{t_{\rm ini}}^tdt'~ \Gamma(t')a^3(t')r^3(t,t').
\label{eq_I}
\end{align}
Here the scale factor $a$ and the comoving coordinate $r(t,t')$ are given by
\begin{align}
a(t)
&=a(t_{\rm ini}) e^{H(t - t_{\rm ini})},
\label{eq_a} \\
r(t,t')
&= \int_{t'}^t dt'' \frac{1}{a(t'')}
= \frac{e^{-H(t' - t_{\rm ini})} - e^{-H(t - t_{\rm ini})}}{a(t_{\rm ini})H},
\label{eq_r}
\end{align}
where the initial time $t_{\rm ini}$ is set to be well before the transition,
and the wall velocity is assumed to be luminal in Eq.~(\ref{eq_r})\footnote{
In our model, the phase transition occurs at quite low temperature ($T/M \ll 1$, 
or $\alpha$ defined in Eq.~(\ref{eq_alpha}) satisfies $\alpha \gg 1$) 
in most of the parameter region.
In such cases, bubble walls are likely to approach the speed of light 
(so-called ``runaway" of bubble walls),
and therefore this assumption is justified.
}.
Substituting Eq.~(\ref{eq_Gamma2}), (\ref{eq_a}) and (\ref{eq_r}) into Eq.~(\ref{eq_I}),
and regarding $S$ as a function of time, one sees that $I$ becomes
\begin{align}
I(t)
&= \frac{4\pi M^4}{3H^4}I'(t),
\label{eq_I2}
\end{align}
with
\begin{align}
I'(t)
&\equiv \int_{\tau_{\rm ini}}^0 d\tau' e^{-S(\tau')}(1-e^{\tau'})^3, \\
\tau_{\rm ini}
&\equiv H(t_{\rm ini} - t) < 0.
\label{eq_Ipr}
\end{align}
Note that we have factored out the $M$ dependence in Eq.~(\ref{eq_I2}),
and therefore $I'$ depends only on $\alpha_{B-L}(0)$.
Also note that, in Eq.~(\ref{eq_Ipr}), the integrand is essentially $\sim e^{-S(\tau')}$ 
for the rescaled time $\tau'$ (or the integration time variable $t'$) different from the endpoint $0$ 
(or endpoint $t$) by a few Hubble times.

\begin{figure}
\begin{center}
\includegraphics[scale=0.85]{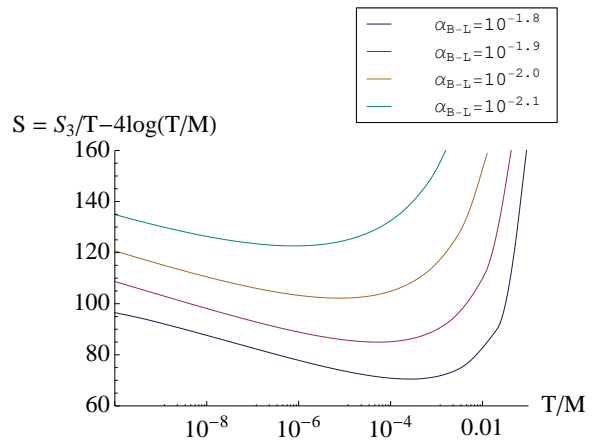}
\caption {\small
Plot of $S$ as a function of $T/M$.
The blue, red, yellow and green lines correspond to 
$\alpha_{B-L} = 10^{-1.8}(\simeq 0.016),~10^{-1.9}(\simeq 0.013),
~10^{-2}(= 0.01)$ and $10^{-2.1}(\simeq 0.008)$, respectively. 
As the temperature decreases, $S$ first drops due to the $S_3/T$ contribution,
while it starts to increase at some point because of the $\ln (T/M)$ contribution.
}
\label{fig_ST}
\end{center}
\end{figure}

Fig.~\ref{fig_ST} is the plot of $S$ as a function of $T/M$.
The blue, red, yellow and green lines correspond to 
$\alpha_{B-L}(0) = 10^{-1.9},~10^{-2.0},~10^{-2.1}$
and $10^{-2.2}$, respectively.
As the temperature decreases, $S$ first drops and then begins to increase.
The former and latter behavior is due to $S_3/T$ term and $\ln (T/M)$ term
in the definition (\ref{eq_S}), respectively.
Since $S$ is exponentiated when calculating $I$, 
the value of $I(t = \infty)$ is mainly determined by the minimum of $S$,
which is quite sensitive to $\alpha_{B-L}(0)$.
This makes the sharp dependence of $I(t  = \infty)$ on $\alpha_{B-L}(0)$ 
in Fig.~\ref{fig_I}, where $I$ is calculated with $M = 10^4$ GeV and 
$\alpha_{B-L}(0) = 10^{-2.12}$, $10^{-2.14}$, $10^{-2.16}$ and $10^{-2.18}$.
This sharp dependence can also be confirmed 
in the behavior of $I'(t=\infty)$ shown in Fig.~\ref{fig_Ipr}.

%%%%%%%%%%%%%%%%%%%%%%%%%%%%%%%%%%%%%%%%%%%%%%%%%%
\subsubsection{Criteria for the phase transition}
%%%%%%%%%%%%%%%%%%%%%%%%%%%%%%%%%%%%%%%%%%%%%%%%%%

\begin{figure}
\begin{center}
\includegraphics[scale=0.85]{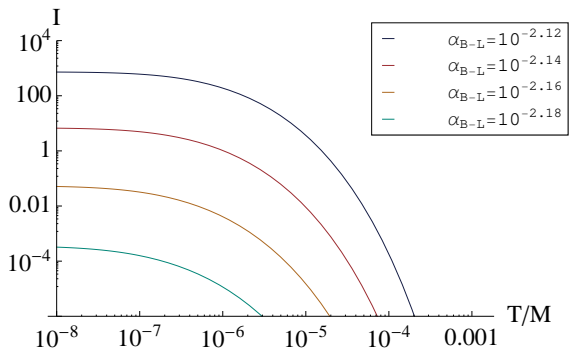}
\caption {\small
Plot of $I$ defined in Eq.~(\ref{eq_I}) for 
$\alpha_{B-L} = 10^{-2.12} \simeq 0.0076$ (blue),
$10^{-2.14} \simeq 0.0072$ (red),
$10^{-2.16} \simeq 0.0069$ (yellow),
$10^{-2.18} \simeq 0.0066$ (green).
$M$ is fixed to be $10^4$GeV.
}
\label{fig_I}
\end{center}
\end{figure}

\begin{figure}
\begin{center}
\includegraphics[scale=0.85]{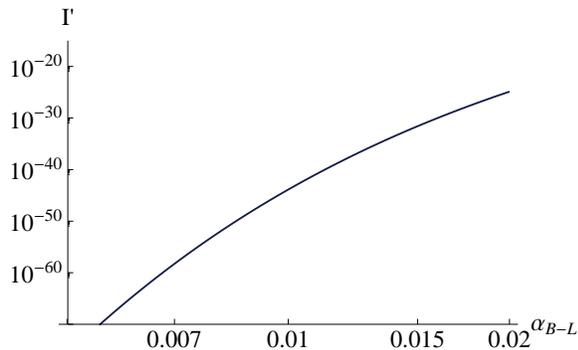}
\caption {\small
$I'(t = \infty)$ as a function of $\alpha_{B-L}$.} 
\label{fig_Ipr}
\end{center}
\end{figure}

We adopt $p(t = \infty)$, or $I(t = \infty)$, as the indicator for the completion of the transition.
The larger $I(t = \infty)$ is, the larger the typical volume of the region without any false vacuum becomes.
Note that, in order to realize the homogeneous CMB spectrum with the $e$-folding $N \sim 50$--$60$ as observed, 
the typical volume of such region must be large enough, say $I(t = \infty) \gtrsim {\mathcal O}(10)$.
In the following discussion we set the condition for the completion of the transition to be 
\begin{align}
I(t = \infty) 
&> I_C,
\;\;\;\;\;\;
I_C = 100,
\label{eq_IC}
\end{align}
and consider the GW production in the regions 
where the phase transition is successfully completed.
Note that the results presented in Sec.~\ref{sec_Results} on $\alpha_{B-L}$--$M$ plane 
show little dependence on the value of $I_C$, 
since $I(t = \infty)$ is quite sensitive to the value of $\alpha_{B-L}$ (see Figs.~\ref{fig_I}--\ref{fig_Ipr}).
Note also that 
there always remain false vacuum regions since the function $I$ is bounded from above in our scenario.
We briefly discuss the fate of the false vacuum regions in Appendix~\ref{app_fate}.

%%%%%%%%%%%%%%%%
%%%%%%%%%%%%%%%%

%%%%%%%%%%%%%%%%
%%%%%%%%%%%%%%%%

%%%%%%%%%%%%%%%%
%%%%%%%%%%%%%%%%

%%%%%%%%%%%%%%%%%%%%%%%%%%%%%%%%%%%%%%%%%%%%%%%%%%
%%%%%%%%%%%%%%%%%%%%%%%%%%%%%%%%%%%%%%%%%%%%%%%%%%
\section{First order phase transition and Gravitational waves}
\label{sec_GW}
%%%%%%%%%%%%%%%%%%%%%%%%%%%%%%%%%%%%%%%%%%%%%%%%%%
%%%%%%%%%%%%%%%%%%%%%%%%%%%%%%%%%%%%%%%%%%%%%%%%%%

In this section we summarize the properties of GWs produced by cosmological first order phase transitions.

When a scalar field is trapped in the false vacuum,
first order phase transitions can occur in association with the tunneling of the scalar field 
from the false vacuum to the true one.
The transition proceeds with the nucleation of bubble seeds,
their expansion, and the collision of the bubbles and subsequent thermalization.
Though a single spherical bubble cannot produce tensor modes in the energy-momentum tensor and hence GWs,
the collision of many bubbles violates the spherical symmetry and as a result GWs are produced.

In first order phase transitions,
the sources of GWs are classified into the following~\cite{Caprini:2015zlo}: 
bubble wall collisions~\cite{Kosowsky:1991ua,Kosowsky:1992rz,Kosowsky:1992vn,Kamionkowski:1993fg,Caprini:2007xq,Huber:2008hg}, 
turbulence~\cite{Dolgov:2002ra,Nicolis:2003tg,Caprini:2006jb,Gogoberidze:2007an,Kahniashvili:2008er,Kahniashvili:2008pe,Megevand:2008mg,Kahniashvili:2009mf,Caprini:2009yp,Caprini:2010xv}
and sound waves~\cite{Hindmarsh:2013xza,Hindmarsh:2015qta,Kalaydzhyan:2014wca}.
The first occurs due to the scalar field configuration, 
while the last two originate from the dynamics in the thermal plasma.
In our setup, the transition typically occurs after the scalar field dominates 
the energy density of the universe, and therefore we focus on bubble wall collisions 
as the source of GWs in the following.
However, one should note that a nonnegligible amount of GWs might be radiated from 
turbulence and sound waves after the phase transition.
Therefore our results should not be regarded as giving the exact estimation of 
the amplitude nor the shape of the GW spectrum, but as a lower limit on the GW production.

%%%%%%%%%%%%%%%%%%%%%%%%%%%%%%%%%%%%%%%%%%%%%%%%%%
\subsection{Gravitational wave spectrum}
%%%%%%%%%%%%%%%%%%%%%%%%%%%%%%%%%%%%%%%%%%%%%%%%%%

In the literature, 
GW spectrum from first order phase transition is often parameterized by two parameters $\alpha$ and $\beta$.
The former is the ratio of the released energy density $\epsilon_*$ 
to radiation energy density at the transition
\begin{align}
\alpha
&= \frac{\epsilon_*}{\frac{\pi^2}{30}g_*' T_N^4},
\label{eq_alpha}
\end{align}
where $T_N$ and $g_*' = 116$ denote 
the temperature and the number of relativistic degrees of freedom in the thermal bath just before the transition.
The other parameter $\beta$ is defined by the nucleation rate per unit volume as
\begin{align}
\Gamma
&= \Gamma_0 e^{\beta t}.
\end{align}
We explain how to estimate $\alpha$ and $\beta$ from the scalar potential in the next subsection.

In this paper, we employ the GW spectrum presented in Ref.~\cite{Huber:2008hg},
where GW spectrum from many bubble collisions is numerically calculated.
In their calculation so-called envelope approximation~\cite{Kosowsky:1991ua,Kosowsky:1992rz,Kosowsky:1992vn,Kamionkowski:1993fg} is adopted,
in which the energy of collided bubble walls is assumed to be instantly transformed into radiation,
and only uncollided bubble walls are taken into account as the source of GWs.
Even though this approximation does not give the full GW spectrum,
we expect that their result gives at least a lower bound for the GW spectrum from bubble collisions.
Also, 
The peak frequency $f_{\rm peak}$ and the GW amplitude at the peak frequency 
$\Omega_{\rm GW,peak} \equiv \Omega_{\rm GW}(f_{\rm peak})$ are estimated as~\cite{Huber:2008hg}
\begin{align}
&f_{\rm peak}
\simeq 17 \; \left(\frac{f_*}{\beta}\right) \left(\frac{\beta}{H_\ast}\right)
\left(\frac{T_{\ast}}{10^8~\text{GeV}}\right)\left(\frac{g_\ast}{100}\right)^{\frac{1}{6}}~[\text{Hz}],
\label{eq_fPeak} \\
&h_0^2\Omega_{\rm GW, peak} \nonumber \\
&\simeq 1.7 \times 10^{-5}
\kappa^2 \Delta \left( \frac{\beta}{H_*} \right)^{-2}
\left(\frac{\alpha}{1+\alpha}\right)^2 
\left( \frac{g_*}{100} \right)^{-\frac{1}{3}},
\label{eq_h2OmegaPeak}
\end{align}
where $H_*$, $T_{*}$ and $g_* = 106.75$ are the Hubble parameter, the temperature and 
the effective degrees of freedom in the thermal bath
after the phase transition, respectively\footnote{
Here, we assume that the system becomes in thermal equilibrium at least within a few Hubble time after the phase transition.
}.
Also, $h_0$ denotes the reduced Hubble constant at present.
Other parameters $\Delta$ and $f_*/\beta$ are given by
\begin{align}
\Delta
&= \frac{0.11v_b^3}{0.42 + v_b^2}, \\
\frac{f_*}{\beta}
&= \frac{0.62}{1.8 - 0.1v_b + v_b^2},
\end{align}
where $v_b$ is the velocity of the energy front of the bubbles. 
In addition, the efficiency $\kappa$ is defined as 
the fraction of the released energy density $\epsilon_*$ 
localized around the bubble walls.
Note that we include the energy stored in the form of the scalar field 
as well as that in the fluid in our definition of $\kappa$.
Though $\kappa$ depends separately on $\alpha$ and $v_b$ in general,
we take the expression for $\kappa$ with so-called Jouguet detonation 
as a benchmark~\cite{Kamionkowski:1993fg}
\begin{align}
\kappa
&= \frac{1}{1+0.715\alpha} \left( 0.715\alpha + \frac{4}{27}\sqrt{\frac{3\alpha}{2}} \right),
\label{eq_kappa}
\end{align} 
and that for $v_b$ under the same assumption~\cite{Steinhardt:1981ct}
\begin{align}
v_b
&= \frac{1/\sqrt{3} + \sqrt{\alpha^2 + 2\alpha/3}}{1+\alpha}.
\label{eq_vb}
\end{align}
Note that, in the present model, $\alpha$ is typically much larger than unity in the parameter region we are interested in,
and in such cases bubble walls are likely to runaway 
rather than expand with detonation~\cite{Espinosa:2010hh,Bodeker:2009qy}.
However, since in runaway cases most of the released energy is likely to be 
localized at the bubble walls expanding with a velocity close to the speed of light,
we expect that $\kappa \to 1$ and $v_b \to 1$ are realized for $\alpha \gg 1$.
Thus Eqs.~(\ref{eq_kappa})--(\ref{eq_vb}) are expected to give a good estimate for produced GWs even in such a case.
Therefore we use Eqs.~(\ref{eq_fPeak})--(\ref{eq_vb}) in the following analysis.

For the frequency dependence of the GW spectrum, we follow the result in Ref.~\cite{Huber:2008hg}
and approximate it as
\begin{align}
\Omega_{\rm GW}
&= 
\left\{
\begin{matrix}
\Omega_{\rm GW, peak} (f/f_{\rm peak})^3 
&\;\;\; (f < f_{\rm peak}) \\
\Omega_{\rm GW, peak} (f/f_{\rm peak})^{-1}
&\;\;\; (f > f_{\rm peak}) \\
\end{matrix}
\right. .
\label{eq_OmegaSignal}
\end{align}
%%

%%%%%%%%%%%%%%%%%%%%%%%%%%%%%%%%%%%%%%%%%%%%%%%%%%
\subsection{Estimate of the bounce action}
\label{subsec_bounce}
%%%%%%%%%%%%%%%%%%%%%%%%%%%%%%%%%%%%%%%%%%%%%%%%%%

For the parameter values with which the phase transition completes,
the transition occurs at $I \sim 1$ for most of the spacial region.
We define the temperature at the bubble nucleation $T_N$ as the temperature at $I = 1$.
Then $\alpha$, $\beta$ and $T_*$ are obtained in the following way.
First, the nucleation speed $\beta = \dot{\Gamma}/\Gamma$ is estimated as\footnote{
Note that this condition differs from the one often used in the literature
\begin{align}
\frac{\beta}{H_*}
&= \left. T \frac{d(S_3/T)}{dT} \right|_{T=T_N},
\end{align}
by $4$, 
because we have taken into account the factor $A \sim T^4$ in Eq.~(\ref{eq_Gamma1}).
}
\begin{align}
\frac{\beta}{H_*}
&= \left. T \frac{dS}{dT} \right|_{T=T_N}.
\label{eq_betaH}
\end{align}
The temperature just after the transition $T_*$ is estimated from the Friedmann equation as
\begin{align}
3M_P^2H_*^2
&= \frac{\pi^2}{30}g_*T_*^4,
\end{align}
with $M_P\simeq 2.4\times 10^{18}$ GeV denoting the reduced Planck mass.
The released energy density $\epsilon_*$ is estimated from the thermodynamic relation~\cite{Espinosa:2008kw}
\begin{align}
\epsilon_*
&= \left[ - \Delta V_{\rm min}(T) + T\frac{d}{dT} \Delta V_{\rm min}(T) \right]_{T=T_N},
\end{align}
where $\Delta V_{\rm min}$ is the temperature-dependent minimum of the effective potential
$\Delta V_{\rm eff} \equiv V_{\rm eff}(\phi,T) - V_{\rm eff}(0,T)$.

%%%%%%%%%%%%%%%%%%%%%%%%%%%%%%%%%%%%%%%%%%%%%%%%%%
\subsection{Typical behavior of GW parameters in classically conformal models}
\label{subsec_typical}
%%%%%%%%%%%%%%%%%%%%%%%%%%%%%%%%%%%%%%%%%%%%%%%%%%

Here we briefly discuss the typical behavior 
of the parameters which determine the properties of GW spectrum
in the present model.
For sufficiently low temperature $T \ll M$,
the effective potential around the origin $\phi \lesssim T$ is 
approximately given by 
the following quadratic and negative quartic terms (see Eq.~(\ref{eq_Veff_approx}))
\begin{align}
V_{\rm eff}
&\simeq \frac{g_{B-L}^2(t)}{2}T^2\phi^2+\frac{\lambda_{\rm eff}(t)}{4}\phi^4,\\
t
&=\ln (T/M).
\end{align}
In such a case, the exponent in the nucleation rate is estimated as~\cite{Linde:1981zj}
\begin{align}
S&=\frac{S_3}{T}-4\ln(T/M),\\
\frac{S_3}{T}
&\simeq  - 9.45 \times \frac{g_{B-L}(t)}{\lambda_{\rm eff}(t)}.
\label{eq_S3Testimation}
\end{align}
Note that $S_3/T$ depends on $T$ only through the running of the coupling $t = \ln (T/M)$,
which makes the dependence of $S$ on $T$ small and as a result $\beta/H$ small ($\sim {\mathcal O}(1)$).
In addition, ultra-supercooling $\alpha \sim M^4 / T_N^4 \gg 1$ is expected 
since the transition occurs not at $T_N \sim M$ but $T_N \ll M$.
This is because the temperature has to decrease by some orders of magnitude from $M$
in order for $S$ to decrease sufficiently (by $\sim {\mathcal O}(10)$) to trigger the transition (see Fig.~\ref{fig_ST})
due to the very weak dependence of $S$ on $T$.

As one can see from Eq.~(\ref{eq_h2OmegaPeak}),
large $\alpha$ and small $\beta/H$ are required to make 
the amplitude of GWs larger.
Therefore, a large amplitude of GWs is expected in the classically conformal $B-L$ model.
This property seems to be universal in models with classical conformal invariance
as mentioned in Ref.~\cite{Konstandin:2011dr},
since above discussion depends only on the weak dependence of 
the nucleation rate to the temperature.

%%%%%%%%%%%%%%%%%%%%%%%%%%%%%%%%%%%%%%%%%%%%%%%%%%
%%%%%%%%%%%%%%%%%%%%%%%%%%%%%%%%%%%%%%%%%%%%%%%%%%
\section{Results}
\label{sec_Results}
%%%%%%%%%%%%%%%%%%%%%%%%%%%%%%%%%%%%%%%%%%%%%%%%%%
%%%%%%%%%%%%%%%%%%%%%%%%%%%%%%%%%%%%%%%%%%%%%%%%%%

In the following we show the contours of the quantities related to the GW spectrum 
on the $M$--$\alpha_{B-L}$ plane,
where $\alpha_{B-L}$ is the shorthand notation for $\alpha_{B-L} (0)$.
Before focusing on each figure, we mention the red, green and yellow lines 
common in Figs.~\ref{fig_alpha}--\ref{fig_ObservableRegion}:
\begin{itemize}
\item
Red line : 
Above this line, the coupling $\alpha_{B-L}$ blows up below the Planck scale (see Eq.~(\ref{eq_aBLSolution})).
\item
Green line : 
The region left to this line is excluded by $Z'$ search,
as explained in Sec.~\ref{sec_setup}.
\item
Yellow line : 
Below this line, the transition condition (\ref{eq_IC}) is not satisfied.
\end{itemize}
We consider the parameter region which avoids these constraints below.

%%%%%%%%%%%%%%%%%%%%%%%%%%%%%%%%%%%%%%%%%%%%%%%%%%
\subsection{Numerical results for GW amplitude}
%%%%%%%%%%%%%%%%%%%%%%%%%%%%%%%%%%%%%%%%%%%%%%%%%%

\begin{figure}
\begin{center}
\includegraphics[scale=0.85]{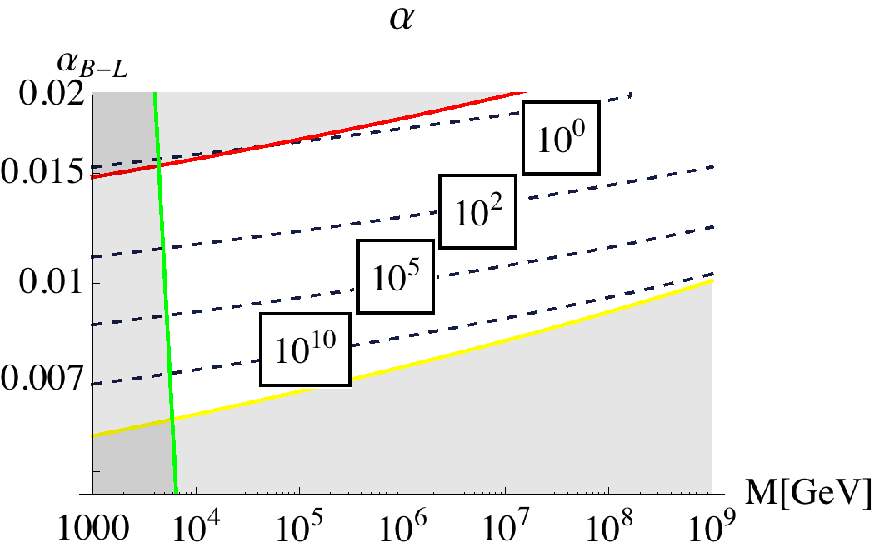}
\caption {\small
Contours of $\alpha$.
The contours correspond to $\alpha = 10^0$, $10^2$, $10^5$ and $10^{10}$
from top to bottom.
Red, green and yellow lines are explained at the beginning of Sec.\ref{sec_Results}.
}
\label{fig_alpha}
\end{center}
\begin{center}
\includegraphics[scale=0.85]{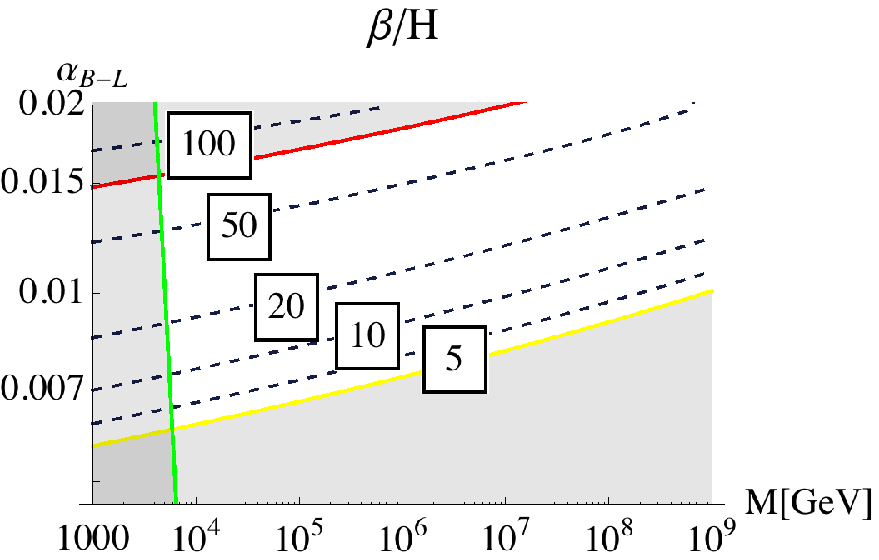}
\caption {\small
Contours of $\beta/H$.
The contours correspond to $\beta / H = 100$, $50$, $20$, $10$ and $5$
from top to bottom.
}
\label{fig_beta}
\end{center}
\end{figure}

We first show the contours of $\alpha$ and $\beta$
in Figs.~\ref{fig_alpha}--\ref{fig_beta}.
In these figures, one sees that
the ultra-supercooling $\alpha \gg 1$ and
the small $\beta/H$ ($\sim {\mathcal O}(1)$--${\mathcal O}(10)$)
mentioned in Sec.~\ref{subsec_typical}
are realized\footnote{
In the parameter region with $\beta/H \sim 1$,
the effect of de Sitter expansion during GW production may not be neglected, 
and the predictions on $f_{\rm peak}$ and $\Omega_{\rm GW, peak}$ may change by some factor.
}.
With a fixed value of $M$, 
smaller $\alpha_{B-L}$ means larger $\alpha$ and smaller $\beta/H$.
Both of these result from the weaker running of couplings for smaller $\alpha_{B-L}$: 
the weak running delays the transition time ($S \sim 100$ in Fig.~\ref{fig_ST}),
and also makes the derivative of $S_3/T$ smaller.

%%%%%%%%%%%%%%%%
%%%%%%%%%%%%%%%%

\begin{figure}
\begin{center}
\includegraphics[scale=0.85]{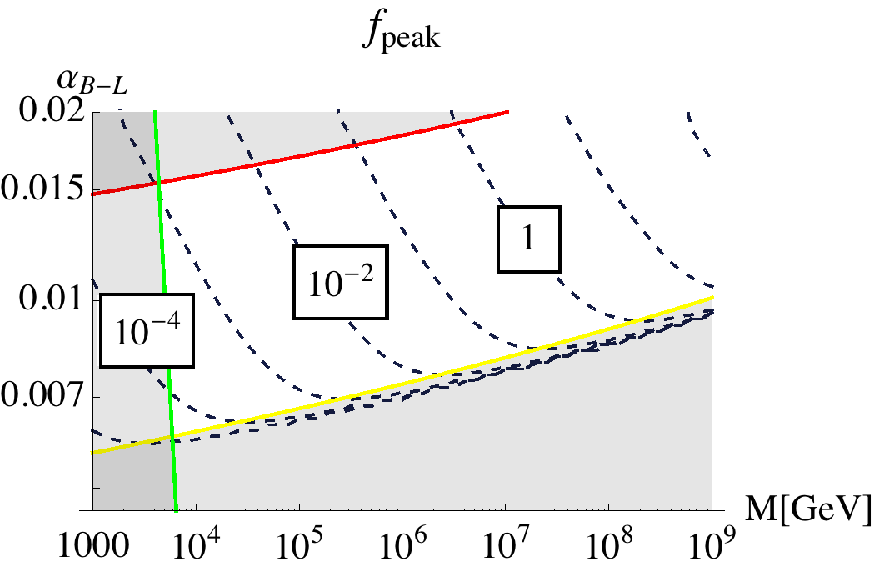}
\caption {\small
Contours of the peak frequency $f_{\rm peak}$
}
\label{fig_fPeak}
\end{center}
\begin{center}
\includegraphics[scale=0.85]{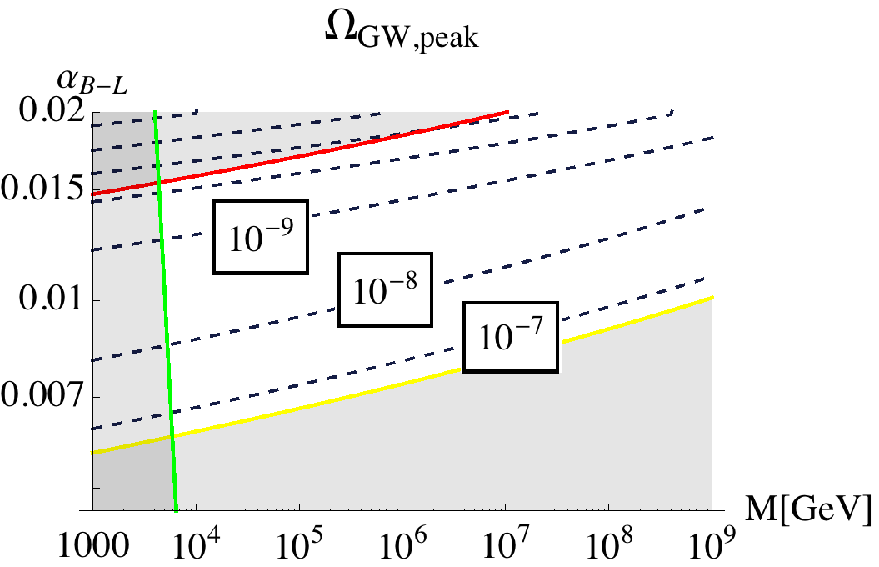}
\caption {\small
Contours of GW amplitude $\Omega_{\rm GW, peak}$ at the peak frequency.
}
\label{fig_OmegaPeak}
\end{center}
\end{figure}

Once we obtain $\alpha$ and $\beta$, 
we can calculate the peak frequency and amplitude of the GW spectrum
using Eqs.~(\ref{eq_fPeak}) and (\ref{eq_h2OmegaPeak}).
Figs.~\ref{fig_fPeak}--\ref{fig_OmegaPeak} show the contours of  
$f_{\rm peak}$ and $\Omega_{\rm GW,peak}$, respectively.
The peak frequency covers $\sim 0.01$--$1$Hz, 
the typical frequency band searched by future space interferometers,
while the GW amplitude becomes as large as $\Omega_{\rm GW, peak} \sim 10^{-7}$--$10^{-9}$ in such a region.

%%%%%%%%%%%%%%%%
%%%%%%%%%%%%%%%%

\begin{figure}
\begin{center}
\includegraphics[scale=0.85]{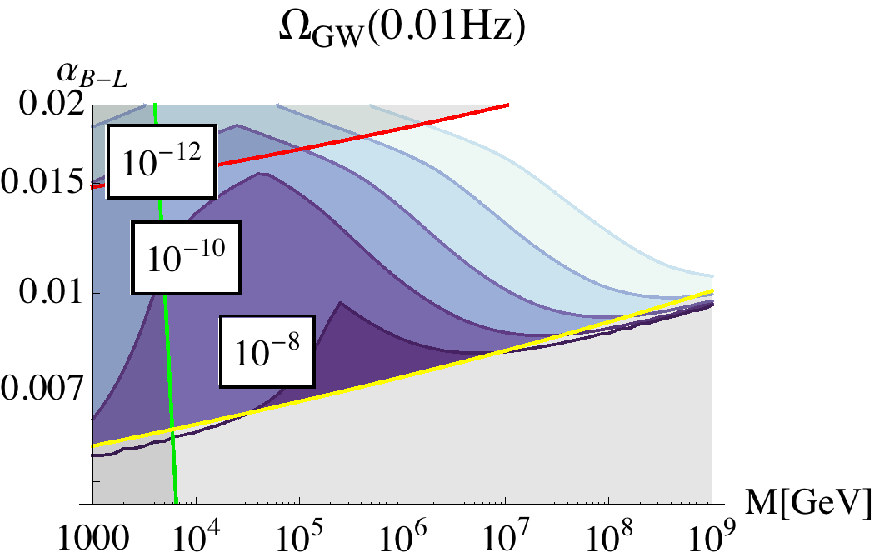}
\caption {\small
Contours of GW amplitude $\Omega_{\rm GW}$ at $f=0.01$ Hz.
}
\label{fig_Omega0.01Hz}
\end{center}
\begin{center}
\includegraphics[scale=0.85]{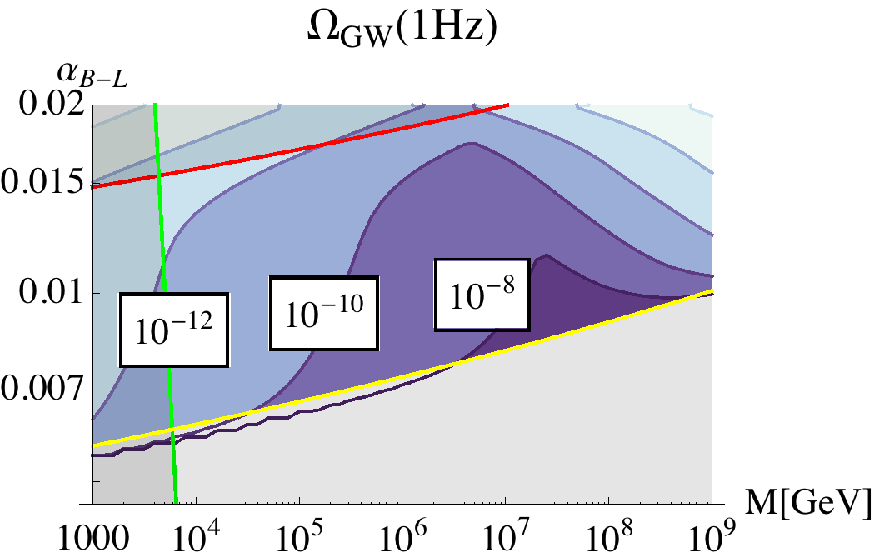}
\caption {\small
Contours of GW amplitude $\Omega_{\rm GW}$ at $f=1$ Hz.
}
\label{fig_Omega1Hz}
\end{center}
\end{figure}

Figs.~\ref{fig_Omega0.01Hz}--\ref{fig_Omega1Hz} show the GW amplitude at $f = 0.01$Hz and $1$Hz.
In both frequency bands, GW amplitude as large as $\sim 10^{-8}$ is realized for some parameter values.
At the same time, 
the GW amplitude $\Omega_{\rm GW} \sim 10^{-8}$--$10^{-12}$
is realized in a vast parameter region in $10^3{\rm GeV} < M < 10^9{\rm GeV}$.

%%%%%%%%%%%%%%%%
%%%%%%%%%%%%%%%%

%%%%%%%%%%%%%%%%%%%%%%%%%%%%%%%%%%%%%%%%%%%%%%%%%%
\subsection{Detectability}
%%%%%%%%%%%%%%%%%%%%%%%%%%%%%%%%%%%%%%%%%%%%%%%%%%

\begin{figure}
\begin{center}
\includegraphics[scale=0.9]{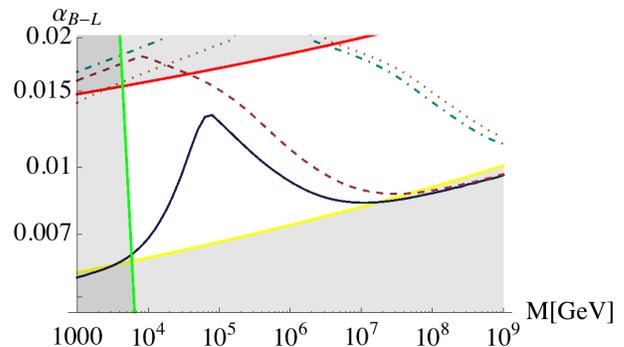}
\caption {\small
Parameter regions where the condition (\ref{eq_DetectableCond}) is satisfied.
The region below the blue-solid, red-dashed, yellow-dotted and green-dot-dashed 
lines satisfy Eq.~(\ref{eq_DetectableCond})
for eLISA, LISA, DECIGO and BBO, respectively.
}
\label{fig_ObservableRegion}
\end{center}
\end{figure}

Finally let us discuss the detectability of the GW spectrum 
realized in this model
by future interferometer experiments.
For simplicity, we approximate the detector noise to be
the radiation pressure noise (for $f < f_{\rm best}$) 
and the shot noise (for $f > f_{\rm best}$)~\cite{Maggiore:1900zz}
\begin{align}
\Omega_{\rm GW}^{(n)}
&= 
\left\{
\begin{matrix}
\Omega_{\rm GW,best}^{(n)} (f/f_{\rm best})^{-1} 
&\;\;\; (f < f_{\rm best}) \\
\Omega_{\rm GW,best}^{(n)} (f/f_{\rm best})^3
&\;\;\; (f > f_{\rm best})
\end{matrix}
\right. ,
\label{eq_OmegaNoise}
\end{align}
and show in Fig.~\ref{fig_ObservableRegion} 
the parameter region which satisfies the condition\footnote{
Note that, with the frequency dependence in Eqs.~(\ref{eq_OmegaSignal}) and (\ref{eq_OmegaNoise}),
this condition is equivalent to 
$\Omega_{\rm GW}(f_{\rm peak}) > \Omega_{\rm GW}^{(n)}(f_{\rm peak})$, or 
$\Omega_{\rm GW}(f_{\rm best}) > \Omega_{\rm GW}^{(n)}(f_{\rm best})$.
}
\begin{align}
\Omega_{\rm GW}(f)
&> \Omega_{\rm GW}^{(n)}(f)
\;\;\;
{\rm for}
\;\;\;
{}^\exists f.
\label{eq_DetectableCond}
\end{align}
The parameter values assumed for $f_{\rm best}$ and $\Omega_{\rm GW,best}^{(n)}$ 
are summarized in Table~\ref{tbl_DetectorSens},
which are estimated from Ref.~\cite{Moore:2014lga} as typical values.
The spectral noise density $S_n$ summarized in the same table is related to $\Omega_{\rm GW}^{(n)}$ as
\begin{align}
\Omega_{\rm GW}^{(n)}(f)
&= \frac{4\pi^2}{3H_0^2}f^3S_n(f),
\end{align}
where $H_0$ is the Hubble parameter at present.

In Fig.~\ref{fig_ObservableRegion}, the region below the dashed lines satisfies
the condition (\ref{eq_DetectableCond})
for eLISA (blue), LISA (red), DECIGO (yellow) and BBO (green), respectively.
An interesting parameter region $M \sim 100$ TeV can be tested by eLISA, 
while a wider region can be explored by the other experiments\footnote{
Better (lower) $\Omega_{\rm GW}^{(n)}(f_{\rm best})$ does not necessarily mean a wider parameter region, 
because the detectability depends also on $f_{\rm best}$.
}.
Furthermore, the parameter region around $M \sim 10^4$ GeV and $\alpha_{B-L} \sim 0.007$
may be searched by SKA~\cite{SKA}, 
since $f_{\rm peak} \lesssim 10^{-4}$Hz and $\Omega_{\rm GW, peak} \sim 10^{-7}$--$10^{-8}$ is realized 
in this region~\cite{Kikuta:2014eja}.

\begin{figure}
\begin{center}
\includegraphics[scale=0.9]{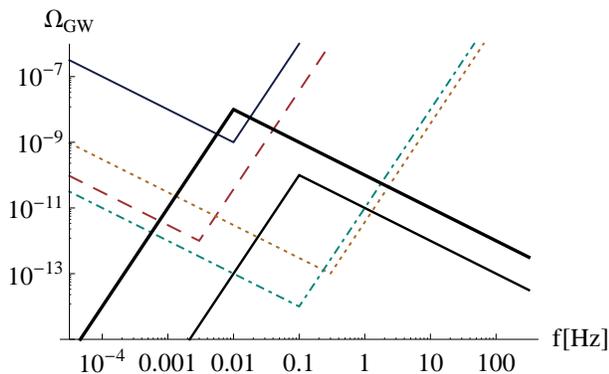}
\caption {\small
Plot of detector sensitivities and GW signals.
Blue-solid, red-dashed, yellow-dotted and green-dot-dashed lines correspond to
eLISA, LISA, DECIGO and BBO, respectively.
Black lines show gravitational signals with peak frequency and amplitude 
$(f_{\rm peak},\Omega_{\rm GW}) = (0.01 {\rm Hz},10^{-8})$ (thick) and $(0.1{\rm Hz},10^{-10})$ (thin). 
}
\label{fig_ObservationImage}
\end{center}
\end{figure}

Fig.~\ref{fig_ObservationImage} is the plot of the GW signal (\ref{eq_OmegaSignal})
for $(f_{\rm peak},\Omega_{\rm GW}) = (0.01 {\rm Hz},10^{-8})$ (black-dashed) 
and $(0.1{\rm Hz},10^{-10})$ (black-dotted), 
and the detector sensitivity (\ref{eq_OmegaNoise})
for eLISA (blue), LISA (red), DECIGO (yellow) and BBO (green), respectively.
These parameter values correspond to 
$(M, \alpha_{B-L}) = (2.4 \times 10^5 {\rm GeV}, 0.010)$ (black-dashed) 
and $(4.5 \times 10^5 {\rm GeV}, 0.017)$ (black-dotted),
respectively.

%%%%%%%%%%
\begin{table}[t]
\begin{tabular}{|l||c|c|c|}
\hline
&$f_{\rm best}$[Hz]&$\Omega_{\rm GW}^{(n)}(f_{\rm best})$&$S_n^{1/2}(f_{\rm best})$[Hz${}^{-1/2}$] \\ \hline\hline
eLISA&$0.01$&$10^{-9}$&$1.9 \times 10^{-20}$ \\ \hline
LISA&$0.003$&$10^{-12}$&$3.7 \times 10^{-21}$ \\ \hline
DECIGO&$0.3$&$10^{-13}$&$1.2 \times 10^{-24}$ \\ \hline
BBO&$0.1$&$10^{-14}$&$1.9 \times 10^{-24}$ \\ \hline
\end{tabular}
\caption{
Detector sensitivities adopted in the analysis.
We included radiation pressure noise and shot noise only,
and scaled the sensitivity as $\Omega_{\rm GW} \propto f^{-1}$ ($f^3$) for 
$f < f_{\rm best}$ ($f > f_{\rm best}$).
}
\label{tbl_DetectorSens}
\end{table}
%%%%%%%%%%

%%%%%%%%%%%%%%%%
%%%%%%%%%%%%%%%%

%%%%%%%%%%%%%%%%
%%%%%%%%%%%%%%%%

%%%%%%%%%%%%%%%%%%%%%%%%%%%%%%%%%%%%%%%%%%%%%%%%%%
\section{Conclusions}
\label{sec_conclusions}
%%%%%%%%%%%%%%%%%%%%%%%%%%%%%%%%%%%%%%%%%%%%%%%%%%

In this paper, 
we have studied the cosmological histories realized 
in the classically conformal $B-L$ gauge extension of the standard model,
and discussed the possibility of probing it using gravitational waves.
Because of the classical conformal invariance of the model,
the thermal trap persists at the origin of the effective potential.
This makes the ultra-supercooling $T \ll M$ possible,
where $M$ is the typical symmetry breaking scale.
In addition, 
the dependence of the nucleation rate to the temperature is suppressed
because of the classical conformal invariance,
which makes the bubbles produced in the transition 
and the resulting GW amplitude larger.
As a result, requiring 
(1) U$(1)_{B-L}$ is broken after the reheating, 
(2) the $B-L$ gauge coupling does not blow up below the Planck scale, 
(3) the thermal phase transition completes, 
the gravitational wave amplitude can be as large as 
$\Omega_{\rm GW} \sim 10^{-8}$ for some model parameters,
and a vast parameter region can be tested by future interferometer experiments.
At the same time, our result of the large amount of GW production is supposed to hold for other classically conformal models.

%%%%%%%%%%%%%%%%%%%%%%%%%%%%%%%%%%%%%%%%%%%%%%%%%%
\section*{Acknowledgments}
%%%%%%%%%%%%%%%%%%%%%%%%%%%%%%%%%%%%%%%%%%%%%%%%%%

RJ and MT are thankful to K.~Nakayama for 
comments and discussions on the manuscript.
Both authors are also grateful to K.~Kohri,
who gave helpful comments to improve the manuscript.
They also appreciate the discussion with S. Iso.
The work of RJ and MT is supported 
by JSPS Research Fellowships for Young Scientists.

%%%%%%%%%%%%%%%%%%%%%%%%%%%%%%%%%%%%%%%%%%%%%%%%%%
\appendix
%%%%%%%%%%%%%%%%%%%%%%%%%%%%%%%%%%%%%%%%%%%%%%%%%%

%%%%%%%%%%%%%%%%%%%%%%%%%%%%%%%%%%%%%%%%%%%%%%%%%%
\section{Discussion on the fate of the false vacuum}
\label{app_fate}
%%%%%%%%%%%%%%%%%%%%%%%%%%%%%%%%%%%%%%%%%%%%%%%%%%

In this appendix we discuss the fate of the false vacuum region
which exists with a finite volume even when 
the condition for the thermal phase transition (\ref{eq_IC}) is satisfied.
The reason for the existence of the false vacuum region is that,
due to the behavior of the nucleation rate $\Gamma$, or its exponent $S$, 
a spacial point in the false vacuum is less and less likely to experience the transition 
after $S$ hits the minimum (see Fig.~\ref{fig_ST}).
In such regions $\phi$ continues to be trapped at the origin of the effective potential,
until it comes out by the de Sitter quantum fluctuations after the universe cools down to $T \sim H$.
Even though such false vacuum region is very rare,
we must take into account the exponential expansion of that region.
Especially, the cosmological history realized in the present model
significantly depends on whether eternal inflation occurs or not.

Eternal inflation occurs when de Sitter fluctuation dominates over the classical motion of the field,
$\Delta \phi_Q \gtrsim \Delta \phi_C$, where $\Delta \phi$ is the amount of $\phi$ motion during one Hubble time,
and the labels $Q$ and $C$ stand for quantum and classical, respectively.
Evaluating at $\phi \simeq H$, one has
\begin{align}
\Delta \phi_Q
&\simeq \frac{H}{2\pi}, \\
\Delta \phi_C
&\simeq \frac{\dot{\phi}}{H}  
\simeq \frac{V_{\rm eff}'}{3H^2}
\simeq \frac{\lambda_{{\rm eff},H} H}{3},
\end{align}
where $\lambda_{{\rm eff},H}$ is the value of $\lambda_{\rm eff}(u)$ at $u=\ln (H/M)$.
Because $\lambda_{{\rm eff},H} \lesssim 10^{-2}$ in the parameter region shown in Sec.~\ref{sec_Results},
eternal inflation occurs in the false vacuum.
(The effective potential can be approximated by the hilltop inflation type around the origin,
in which case the condition for the eternal inflation is studied in Ref.~\cite{Barenboim:2016mmw}.)
In those patches where $\phi$ starts classical rolling towards the true vacuum,
no observers are expected to exist.
However,
some patches of the universe where the inflation finishes may produce observers with 
a quite small but finite probability.
(Of course, even if the inflation ends in one Hubble patch,
the isotropic and homogeneous CMB spectrum with $e$-folding $N \sim 50$--$60$ cannot 
be realized in that single patch.
However, there seems to exist a tiny probability 
with which a large number of Hubble patches experience the transition at the same time
and observers are born inside.)
Because eternal inflation produces infinite number of such patches, 
typical observers realized in this model may find 
themselves not in the patch where the thermal phase transition is successfully completed, 
but in the one which experienced an inflating phase at $\phi \sim 0$ and then 
happened to escape.

If one finds this situation problematic,
the setup can be modified so that a typical observer 
lives in the patch where the thermal phase transition is completed.
One way of doing this is to modify the potential so that
the phase transition is completed at all the spacial points,
which is realized for example by introducing a nonminimal coupling $\phi^2R$
with a negative coefficient of ${\mathcal O}(1)$ or larger, 
which induces a negative mass squared $\sim H^2$ at the origin of the potential
and makes $\phi$ roll down to the symmetry breaking minimum at all the spacial points.
Forbidding the production of some inevitable components such as baryon asymmetry or dark matter
in the patches where the phase transition occurs after the eternal inflation phase 
may be another solution.

%%%%%%%%%%%%%%%%%%%%%%%%%%%%%%%%%%%%%%%%%%%%%%%%%%

%%%%%%%%%%%%%%%%%%%%%%%%%%%%%%%%%%%%%%%%%%%%%%%%%%

%%%%%%%%%%%%%%%%%%%%%%%%%%%%%%%%%%%%%%%%%%%%%%%%%%
\end{document}